\documentclass[10pt,conference]{IEEEtran}
\usepackage[utf8]{inputenc}
\usepackage{amssymb}
\usepackage{amsmath}
\usepackage{mathtools}
\usepackage{ellipsis}
\usepackage{textcomp}
\usepackage{cite}
\usepackage{color}
\usepackage{graphicx}
\usepackage[nolist,nohyperlinks]{acronym}
\usepackage{comment}
\usepackage{csquotes}
\usepackage{caption}
\usepackage{subcaption}
\usepackage{balance} 
\usepackage{amsmath}
\usepackage{multirow}
\usepackage{amsfonts}
\usepackage{amssymb}
\usepackage[version=4]{mhchem}
\usepackage{stmaryrd}
\usepackage{hyperref}
\usepackage{booktabs}
\usepackage{tabularx}
\usepackage{graphicx}
\usepackage{pifont}
\usepackage{caption}
\usepackage[normalem]{ulem}

\usepackage[compact]{titlesec}

\newcolumntype{s}{>{\hsize=0.5\hsize\arraybackslash} X}

\title{RAID: In-Network RA Signaling Storm Detection for 5G Open RAN}
\author{\IEEEauthorblockN{Mohamed Rouili, Yang Xiao, Sihang Liu, Raouf Boutaba}

\IEEEauthorblockA{\{mrouili, yang.xiao, sihangliu, rboutaba\}@uwaterloo.ca}

\IEEEauthorblockA{University of Waterloo, Canada} }
\begin{document}

\begin{acronym}
    \acro{RL}{Reinforcement Learning}
    \acro{ML}{Machine Learning}    
    \acro{3GPP}{Third Generation Partnership Project}
    \acro{TS}{Technical Specifications}
    \acro{RAN}{Radio Access Network}
    \acro{OSM}{Open Source MANO}
    \acro{ETSI}{European Telecommunications Standards Institute}
    \acro{VNF}{Virtual Network Function}
    \acro{VNFD}{Virtual Network Function Descriptor}
    \acro{NSD}{Network Service Descriptor}
    \acro{UE}{User Equipment}
    \acro{RLS}{Radio Link Simulation}
    \acro{VIM}{Virtual Infrastructure Manager}
    \acro{KPI}{Key Performance Indicator}
    \acro{QoS}{Quality of Service}
    \acro{GTP}{GPRS Tunneling Protocol}
\end{acronym}

\maketitle

\thispagestyle{plain}
\pagestyle{plain}

\begin{abstract}

The disaggregation and virtualization of 5G Open RAN (O-RAN) introduces new vulnerabilities in the control plane that can greatly impact the quality of service (QoS) of latency-sensitive 5G applications and services. One critical issue is Random Access (RA) signaling storms where, a burst of illegitimate or misbehaving user equipments (UEs) send Radio Resource Control (RRC) connection requests that rapidly saturate a Central Unit’s (CU) processing pipeline. Such storms trigger widespread connection failures within the short contention resolution window defined by 3GPP. Existing detection and mitigation approaches based on near-real-time RAN Intelligent Controller (n-RT RIC) applications cannot guarantee a timely reaction to such attacks as RIC control loops incur tens to hundreds of milliseconds of latency due to the non-deterministic nature of their general purpose processor (GPP) based architectures. This paper presents RAID, an in-network RA signaling storm detection and mitigation system that leverages P4-programmable switch ASICs to enable real-time protection from malicious attacks. RAID embeds a lightweight Random Forest (RF) classifier into a programmable Tofino switch, enabling line-rate flow classification with deterministic microsecond-scale inference delay. By performing ML-based detection directly in the data plane, RAID catches and filters malicious RA requests before they reach and overwhelm the RRC. RAID achieves above 94\% detection accuracy with a fixed per-flow inference delay on the order of 3.4 microseconds, effectively meeting strict O-RAN control-plane deadlines. These improvements are sustained across multiple traffic loads, making RAID a fast and scalable solution for the detection and mitigation of signaling storms in 5G O-RAN.

\end{abstract}


\section{Introduction}

\begin{figure}[ht!]
\includegraphics[width=\linewidth]{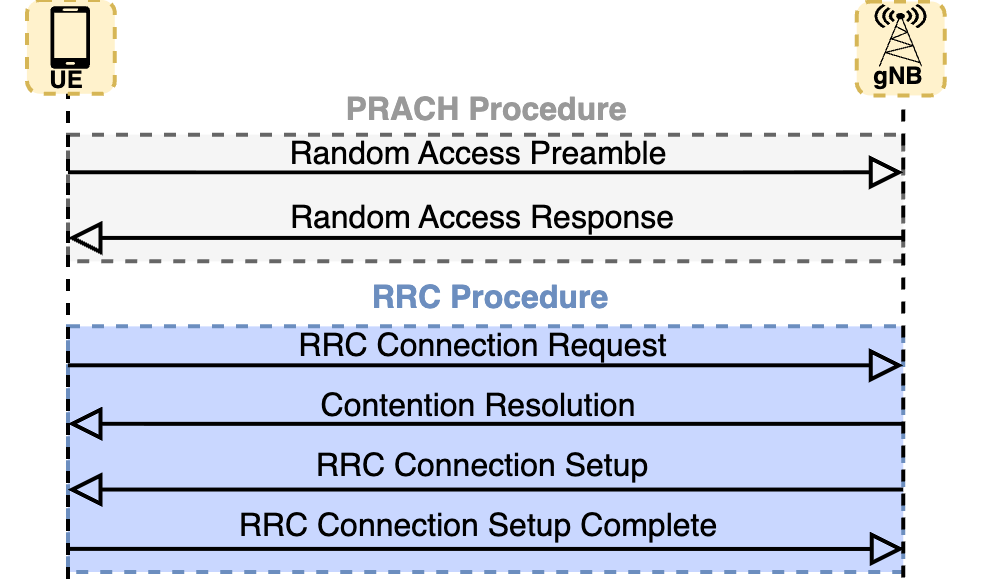}
    \caption{Random Access UE procedure in 5G O-RAN}
    \label{fig:RASS}
\end{figure}

Emerging latency-sensitive applications in 5G and beyond impose increasingly stringent latency and responsiveness requirements on the Open Radio Access Network (O-RAN) control plane. One particularly critical scenario involves the Random Access (RA) procedure, which uses the Radio Resource Control (RRC) in the Centralized Unit (CU) to allow user equipments (UEs) to request a connection to the network. As shown in figure \ref{fig:RASS}, the RA procedure starts with the Physical Random Access Channel (PRACH) exchange. The UE first transmits a random access preamble (MSG1) to acquire uplink timing and a temporary Radio Network Temporary Identifier (RNTI) for subsequent access, the network responds with a random access response (MSG2). After PRACH, the UE issues an RRC Connection Request (MSG3) and the network concludes the exchange with contention resolution (MSG4). The gNB then returns an RRC Connection Setup to which the UE replies with RRC Connection Setup Complete. This five-step RA procedure is standard across several scenarios including UE initial access, re-establishment, or handover requiring uplink synchronization.

The RA procedure is vulnerable to overload due to its contention-based design. A large burst of MSG3 transmissions, either due to misbehaving UEs or deliberate signaling storms can saturate the CU’s RRC processing pipeline and lead to contention resolution failures. These failures delay or deny service to legitimate UEs attempting to access the network. The critical bottleneck in this exchange is the MSG3 to MSG4 phase, which is governed by the Contention Resolution Timer (CRT). This timer is started by a UE after it transmits MSG 3. This timer is configured by the network to indicate how long it waits for successful contention resolution and it ranges from 8 to 64 milliseconds (ms) as standardized by the 3GPP. The timer stops once the contention is resolved and MSG 4 is received. If the CRT expires without resolution, the RA procedure fails, discarding temporary values assigned to UE and informing the upper layers of the failure. In an attack event, attackers can exploit MSG 3. If too many devices repeatedly send MSG 3 to the RAN in rapid bursts with little delay, the CU can exhaust all its computing resources, causing severe Quality of Service (QoS) degradation and potentially a system crash.

This is unacceptable for latency-sensitive deployments, such as ultra-reliable low-latency communication (urLLC) type services and use cases, where any storm detection and mitigation solution must act within single-digit milliseconds to ensure timely detection and handling of attacks before the RA failure is caused by CRT expiration \cite{analysisreportoran}. The O-RAN Alliance \cite{oran-alliance} recognizes signaling storm protection in the RAN as a significant problem and highlights that latency-sensitive urLLC scenarios such as life-critical applications and services are particularly vulnerable. In such cases, even brief disruptions caused by signaling storms can directly impact the life and health of individuals.



Prior RA signaling storm and anomaly detection approaches for the RAN \cite{Hoffmann_2023,Branco_2024,Wen_2024,Neto_2025,Xavier_2023} are built around O-RAN's near-real-time Radio Intelligent Controller (near-RT RIC) architecture. The near-RT RIC introduces near real-time closed-loop intelligence and control in the form of xApps. Even under healthy conditions, these control-loop latencies span 10 ms to 1 s because of xApp and RIC processing on top of E2 interface transport. In addition, similar to CUs and DUs, near-RT RIC implementations primarily rely on general-purpose processors (GPPs) housed in commodity servers. GPPs are not inherently optimized for time-sensitive tasks and high-speed packet processing. This leads to significant overheads, reduced reliability and the inability to guarantee the strict QoS requirements of latency-sensitive use cases such as timely detection and mitigation of RA signaling storm attacks.

In this paper, we introduce \textbf{RAID}, an in-network anomaly detection framework designed to detect and mitigate RA signaling storms in real time. RAID relocates the intelligence traditionally hosted in near-RT RIC xApps directly into the data plane, embedding an ML-based flow classification model within a P4 programmable switch ASIC. RAID is 3GPP and O-RAN compliant and can be integrated within any O-RAN architecture, as it does not require any modifications or special configurations in the RAN. By executing inference directly in the network, RAID bypasses the software and transport bottlenecks of RIC-based systems. This enables real-time detection and mitigation that operates entirely within the strict CRT window and guarantees latency demands of time-sensitive applications. RAID achieves fully deterministic inference latency of just a few microseconds, offering several orders of magnitude faster response compared to RIC-based approaches while maintaining high classification accuracy.

\noindent\textbf{Contributions.} In this work we make the following contributions:

 \begin{itemize}
    \item We propose a 3GPP standard \cite{3gpp} and O-RAN \cite{oran-alliance} compliant in-network system design for line-rate detection and mitigation of RRC signaling storms.
    \item We introduce a flow-based detection and mitigation approach for robust identification of RRC signaling storms.
    \item We implement RAID on a programmable Tofino 1 switch deployed on a real 3GPP and O-RAN compliant 5g lab testbed.
    \item We conduct thorough experimental evaluations demonstrating RAID's performance and scalability.

\end{itemize}

\section{Background and Motivation}\label{sec:background}

\begin{figure}[ht!]
\includegraphics[width=\linewidth]{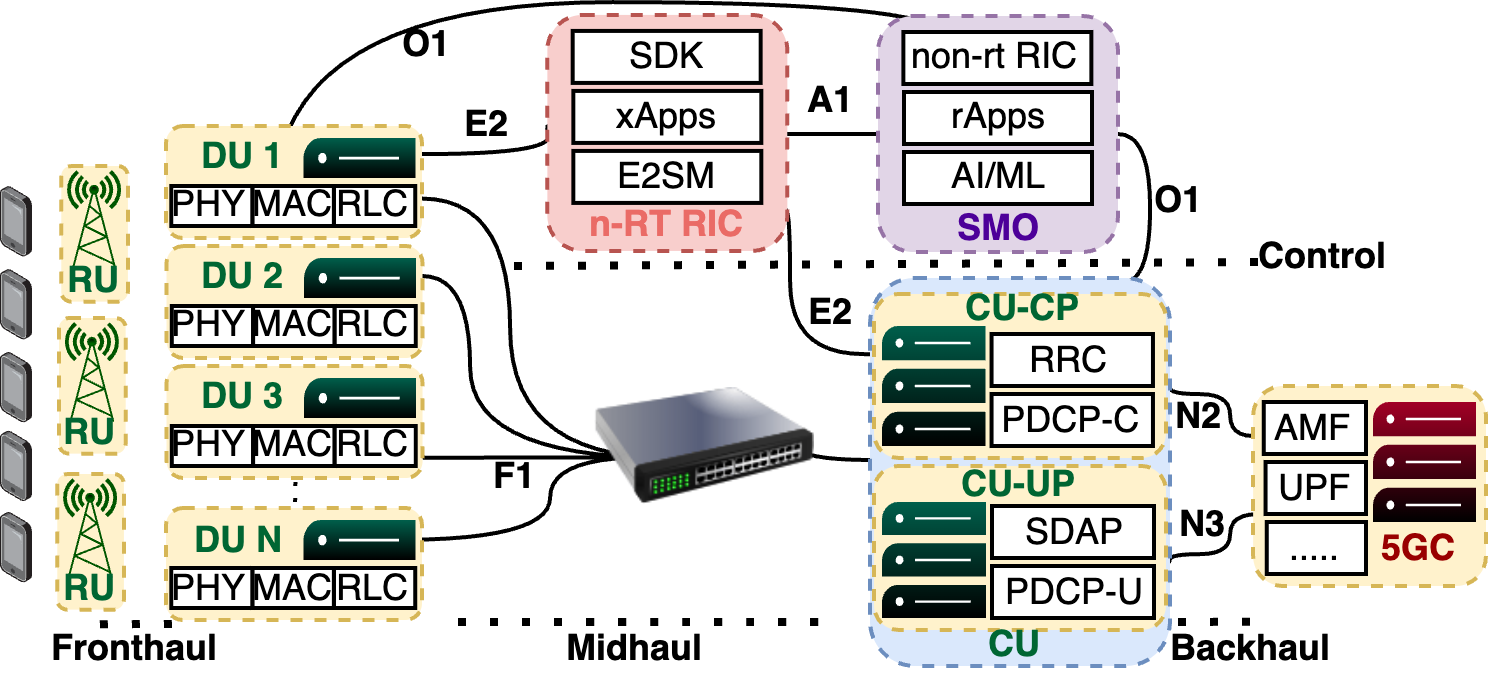}
    \caption{Open Radio Access Network architecture}
    \label{fig:ORAN}
\end{figure}

\subsection{\textbf{O-RAN architecture}}\label{subsec:oranarchitecture}

O-RAN embraces the 3GPP functional split and disaggregates the monolithic base station into logically independent nodes, the Radio Unit (RU), Distributed Unit (DU), and CU, interconnected by open, standardized interfaces as shown in Figure \ref{fig:ORAN}. The RUs represent the radio cell towers deployed at the network edge to handle Layer 1 (L1) physical radio signal transmission from UEs within range. The DUs are responsible for physical resource block allocation and handle key Layer 2 (L2) functions such as Media Access Control (MAC) and Radio Link Control (RLC). The CU serves as the aggregation point of the RAN managing upper-layer functions and serving multiple DUs and UEs. The CU is further disaggregated into control (CU-CP) and user (CU-UP) planes. The CU-CP is responsible for managing the RRC, which involves handling the connections and signaling between UEs and the RAN while CU-UP manages upper user plane functions including the Service Data Adaptation Protocol (SDAP) and Packet Data Convergence Protocol (PDCP). The CU-CP and CU-UP attach to the 5G core's (5GC) control and user plane functions via the N2 and N3 interfaces. Transport in O-RAN is explicitly segmented into fronthaul, midhaul and backhaul. The F1 interface spans the newly introduced midhaul to carry both control and user traffic between DU and CU, enabling flexible placement of these functions across edge and regional sites. Management, orchestration and automation are realized through the Service Management and Orchestration unit (SMO). The non-real-time RIC (non-RT RIC) within the SMO supports rApps for intelligence with slower control loops such as analytics and Artificial Intelligence/Machine Learning (AI/ML) model training. For time-sensitive intelligence such as AI/ML model inference tasks, O-RAN introduced the near-RT RIC. It hosts xApps that interact with the RAN over the standardized E2 interface using E2 Service Models (E2SM). An xApp needs to define E2 Service Models as function-specific protocols on top of the generic E2 Application Protocol (E2AP) to engage with the RAN. For instance, the O-RAN Alliance has demonstrated a few exemplar xApps and E2SMs for use cases like key performance measurements (KPM), RAN slicing management, and traffic steering.

\subsection{\textbf{Limitations of Related Work}}\label{subsec:limapproach}

Existing RA signaling storm and RAN anomaly detection frameworks and approaches  \cite{Hoffmann_2023,Branco_2024,Wen_2024,Neto_2025,Xavier_2023} are built around the O-RAN RIC architecture, typically in the form of a ML-based solution implemented on the near-RT RIC as an xApp. 5G-Spector \cite{Wen_2024} is a state-of-the-art example of such frameworks. It introduces a near-RT RIC xApp solution for diverse Layer 3 (L3) exploits including signaling storms targeting the RRC. Spector’s reported detection and mitigation control loop is in the order of hundreds of milliseconds in practice, which, while acceptable for less urgent anomalies, is too slow to promptly handle fast onset events like RA signaling storms. Recognizing the deficiencies of near-RT RIC approaches, some recent efforts have tried to push intelligence closer to the data plane to shorten the feedback loop. EdgeRIC \cite{KO_2024} introduces a real-time RIC framework co-located at the DU, achieving sub-millisecond control decisions by effectively embedding the control logic at the RAN in the form of µApps. This approach eliminates the network hop to a cloud controller and demonstrates significant improvement in responsiveness. However, EdgeRIC’s scope is limited to L2 functions (PHY/MAC/RLC) and does not support real-time decisions at upper L3 functions such as the RRC. 

Recent works have shown that embedding ML inference directly within programmable hardware enables real-time, line-rate decision making for a variety of tasks  and use cases \cite{Simpson_2022,Zhang_2023,Yan_2024,Akem_2024}. RAID adopts a similar direction through a fully in-network ML design, executing ML inference directly in the data plane to detect and mitigate RA signaling storms at line-rate.

\subsection{\textbf{Key Challenges}}\label{subsec:keychallenges}
To effectively detect and mitigate RRC signaling storms in O-RAN, we identify three key challenges that must be considered.

\begin{figure}[ht!]
\includegraphics[width=\linewidth]{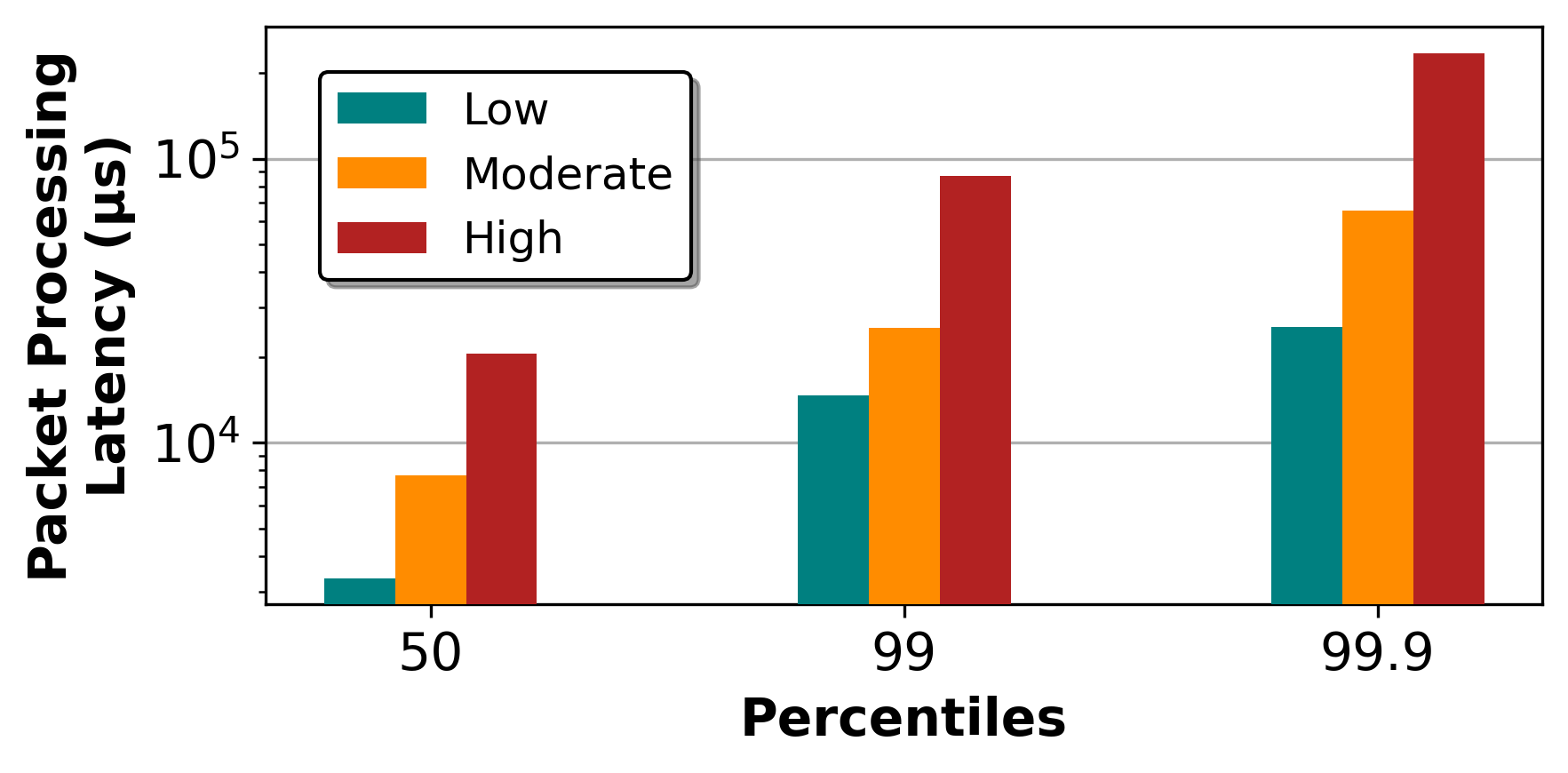}
    \caption{RIC control-loop tail latency }
    \label{fig:xapptail}
\end{figure}

\noindent\textbf{Challenge 1: Meeting real-time detection constraints.} During RA signaling storms, the CRT window becomes the dominant safety margin when it comes to QoS. n-RT RIC control loops running on GPPs plus E2 transport delay cannot guarantee a timely reaction, especially in the tail. As a result, CRT expiry leads to RA failure and blocked service for honest UEs which is unacceptable for latency-critical use cases. We illustrate this through a simple practical experiment on our 5G lab testbed. The experimental setup is discussed in Section \ref{subsec:setup}. We measure the tail packet processing latency of a minimal xApp that parses and prints IP headers of incoming packets before forwarding them to a traffic sink. This setup emulates the usual n-RT RIC control-loop in O-RAN. Figure \ref{fig:xapptail} shows the obtained tail latency results under varrying traffic rates of 10000 packets per second (10K PPS) for low, 30000 for moderate (30K PPS), and 50000 (50K PPS) for high traffic rates. We chose the loads based on the computational limits of our testbed hardware with the higher load indicating near saturation of system resources. Results show a steep degradation in the RIC’s tail performance as load increases. At low loads, delays remain within a few milliseconds, but at moderate and high loads, the 99.9th percentile grows to over $230$~ms several orders of magnitude higher than the median. This sharp rise stems from CPU contention, user-space queuing, and E2 transport overhead inherent to GPP-based RIC control-loops. As a result, even a minimal xApp cannot sustain deterministic timing with reaction latencies far exceeding the 8–60 ms CRT window. 

\noindent\textbf{Challenge 2: Distributed RA storms under CU/DU disaggregation.} Prior signaling storm detection and mitigation schemes and xApps assume a monolithic gNodeB (gNB) where lower-layer congestion is observable before the storm reaches the RRC. In O-RAN, CU/DU disaggregation and virtualization add flexibility and scalability but introduce new storm threat patterns since attackers are able to use distributed RA traffic that appears innocuous at the DU level and only forms a storm when aggregated during F1 transport towards the CU bypassing any detection methods deployed at the lower layers.

\noindent\textbf{Challenge 3: Malicious UE identification.} An important concern lies in the methods used to detect and identify UEs involved in a signaling storm attack. Unlike signaling storm detection at the 5GC where many persistent identifiers such as the International Mobile Subscriber Identity/International Mobile Equipment Identity (IMSI/IMEI) are visible, many identifiers at the RAN such as the RNTI are ephemeral. They change per new establishment or persist only under specific procedures such as a re-establishment, which has been highlighted in the O-RAN Alliance's latest release of its O-RAN use cases analysis report \cite{analysisreportoran}. It discusses alternative detection techniques to identify an aggressive UE that do not rely solely on its identifiers but rather on signal parameters such as signal quality or timing advance. These parameters can help build a unique fingerprint of the aggressive device. However, sophisticated attackers can still get around these detection methods by compromising the chipset itself and spoofing parameters. We argue that a robust storm detection method should adopt an identifier and parameter-agnostic approach that does not rely on any temporary UE identifiers or parameters that can be altered by attackers.

\section{System Design}\label{sec:design}


\begin{figure*}
  \centering
  \begin{subfigure}{0.24\textwidth}
\includegraphics[width=\linewidth]{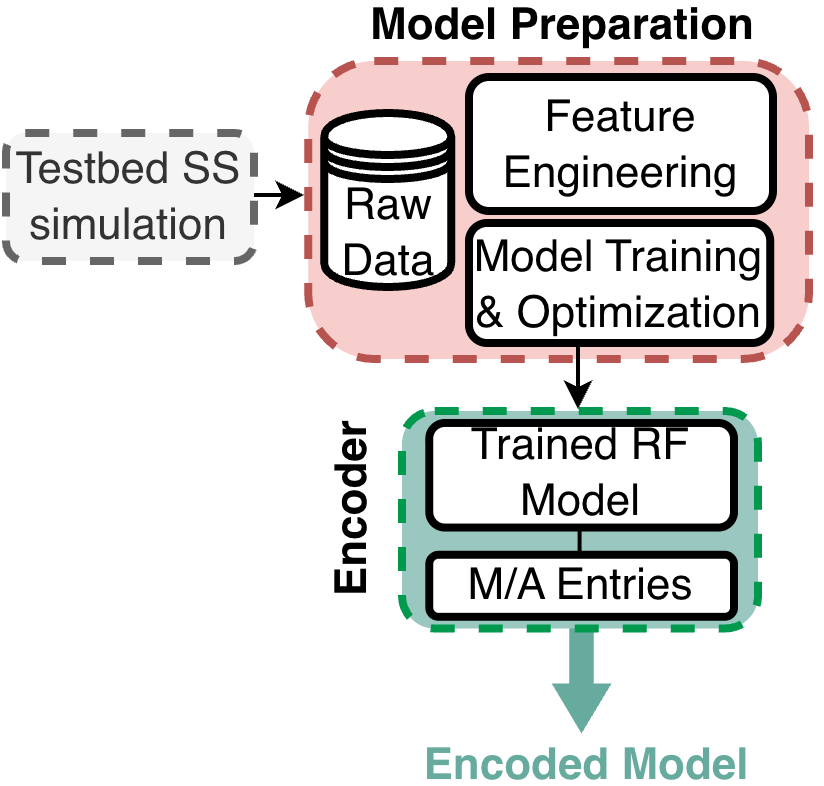}
    \caption{Offline ML preparation}
    \label{fig:offline_raid}
  \end{subfigure}  
  \begin{subfigure}{0.74\textwidth}
\includegraphics[width=\linewidth]{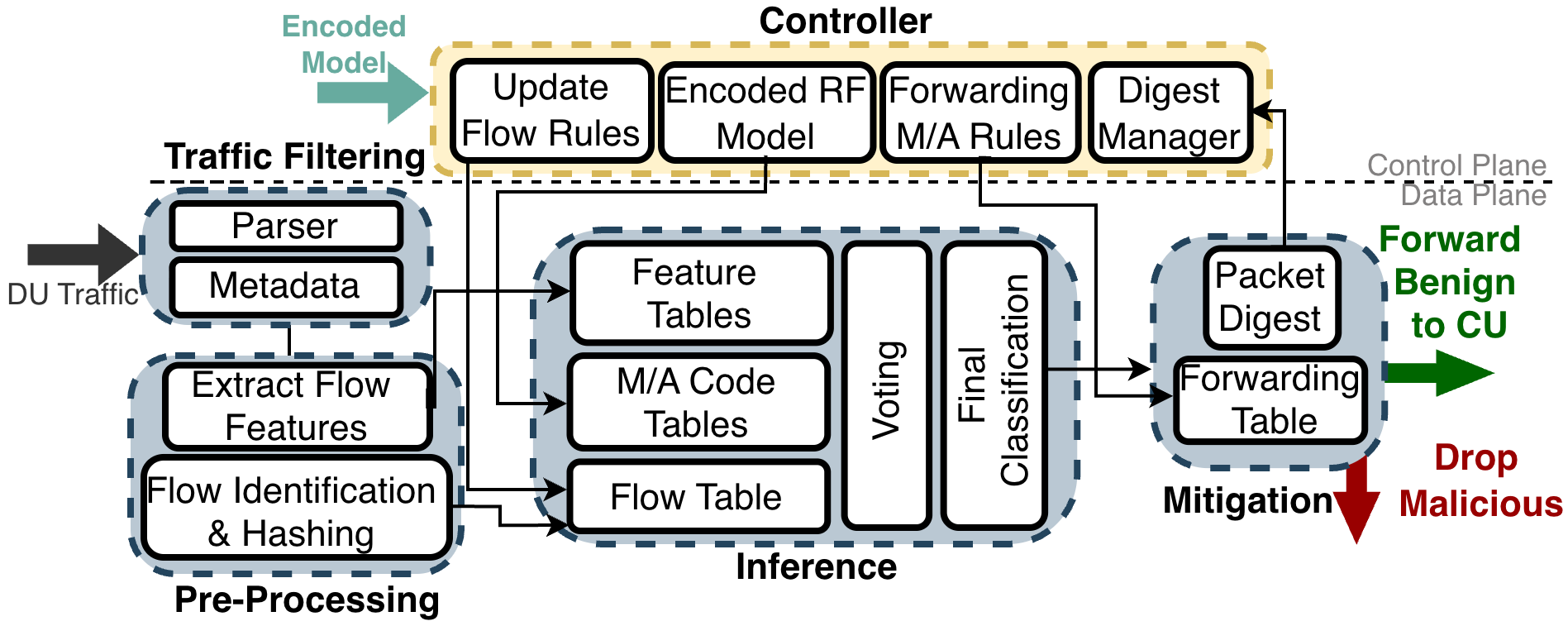}
    \caption{RAID's Architecture Flowchart}
    \label{fig:online_raid}
  \end{subfigure}
\caption{RAID's Architecture}
\end{figure*}

The limitations of xApp-based approaches, existing work, and GPP platforms lead us to design RAID, an in-network system that embeds ML-based RA signaling storm detection and mitigation entirely in the O-RAN data plane on a P4 programmable Tofino 1 switch. Through our system design, we explicitly address the outlined key challenges. Figure \ref{fig:online_raid} shows RAID's architectural components which are divided into data plane and control plane. RAID’s data plane represents its main ML inference P4 pipeline, which is designed to work on an Intel Tofino 1 switch. Programmable-switch ASICs such as Tofino switches are commercially available programmable switches based on the Reconfigurable Match-Action Table (RMT) architecture and the P4 language. Their fixed pipeline stage architecture is optimized for high-speed line-rate packet processing and equipped with equal resources per stage. It contains specialized memory blocks that store packet metadata and maintain match-action table states. RAID's workflow includes offline ML model preparation tasks performed prior to the online deployment as shown in figure \ref{fig:offline_raid}. These tasks include data collection, ML model preparation and the encoding of the trained model into a match/action format to be loaded into the data plane. The controller component is deployed online at runtime and continuously interacts with the data plane by setting up ports, loading the encoded model and managing match action table rules. 

\subsection{\textbf{Design Choices}}\label{sec:designchoices}

Our proposed design addresses the outlined challenges and offers several key benefits as follows.

\noindent \textbf{Line-rate inference.} by operating at line-rate within the switch, RAID eliminates both the control-loop delays of near-RT RIC xApps and packet processing delays on GPP platforms. This enables per-packet microsecond granularity and under $4~\mu$s detection and mitigation to meet strict CRT deadlines. 

\noindent \textbf{In-network integration.} RAID is deployed on the F1 midhaul transport to provide a centralized aggregation point where RA signaling from multiple distributed DUs converges, allowing it to observe aggregate traffic patterns that would otherwise go undetected by lower layers. RAID is O-RAN and 3GPP standard compliant, it sits on the F1 path and integrates seamlessly into the CU/DU disaggregated architecture and does not require any modifications to the standardized Network Functions (NFs) and interfaces.

\noindent \textbf{Flow-based Detection.} To address the unreliability of UE identifiers and parameters, we propose a flow-based detection and mitigation approach. We identify a traffic flow by its 5-tuple: source IP address, destination IP address, source Stream Control Transmission Protocol (SCTP) port, destination SCTP port and protocol number. This approach brings two key benefits to our system design. On one hand, since a flow is identified by multiple persistent packet headers, it prevents attackers from evading detection via spoofed parameters and ensures a robust detection compared to UE-based identifiers in O-RAN. On the other hand, once a flow is classified as malicious or benign, we are able to update table flow rules in RAID's P4 pipeline to allow subsequent packets belonging to that same flow to skip the classification process since their class is already known which reduces processing overhead.



\subsection{System Overview}\label{sec:overview}

The line-rate performance of programmable switch ASICs imposes strict computational constraints: only simple integer and bit-wise operations are supported, no floating-points, and the number of per-packet operations is tightly bounded. Careful ML model selection, feature engineering, and optimization are required to achieve switch compatibility while maintaining traffic classification accuracy comparable to GPP-based deployments. Models must be trained offline and encoded as match-action (M/A) table entries for in-switch inference.

In the context of signaling storm detection, attack behavior is inherently unpredictable as adversaries can arbitrarily vary the rate and timing of RRC connection requests to bypass static detection rules. This lack of consistent or repeatable patterns makes fixed thresholds on RAN parameters ineffective for accurate detection. This motivates the adoption of ML, which offers the flexibility to recognize diverse and evolving attack patterns through data-driven inference rather than manually configured rules.

Within these constraints, Decision Tree (DT) and Random Forest (RF) models emerge as natural candidates for line-rate, in-switch classification. Both have been successfully implemented in data-plane hardware and have demonstrated the feasibility of in-switch ML across a wide range of traffic classification tasks \cite{Zheng_2024,Akem_2024}. Their inference process reduces to threshold comparisons and table lookups, which align well with the Tofino RMT pipeline and enable accurate, low-latency classification at hardware speeds.

\noindent \textbf{Model Preparation.} To the best of our knowledge, there is no publicly available RA signaling storm datasets so we opted to collect our own under realistic conditions. We started by deploying a 3GPP- and O-RAN-compliant 5G lab testbed following the architecture shown in Figure \ref{fig:ORAN} including the Tofino 1 switch placed on the F1 link. We then conducted a set of real-world simulations of signaling storm RA scenarios. Our captured raw data consists of per-flow RRC connection requests (MSG3s), which are precisely the messages adversaries amplify to create signaling storms. From these raw traces we derive switch-feasible, identifier-agnostic features carefully chosen to both capture the temporal patterns that distinguish signaling storms from benign loads while complying with switch constraints. Our six engineered flow features are: packet count, total packet length, max packet length, min packet length, min inter-packet delay (IPD), and max IPD. We partition the dataset into training and testing splits. The main concern for in-network ML inference is the limited stateful memory in the switch so we carefully tune the RF model offline to balance accuracy with switch resource consumption. The final model is passed to the encoder component, which translates the trees into match–action entries to be deployed on the switch.

\noindent \textbf{The Encoder.} Different schemes for mapping RFs into P4 switch pipelines have been previously proposed in the literature \cite{Busse_2019,Xie_2022,Zheng_2024}. Among these schemes, we base our model encoding on Planter \cite{Zheng_2024}, a state-of-the-art RF encoding scheme that has been successfully adopted by several other works. Planter introduces a model decomposition strategy that separates inference into two types of M/A tables: feature tables and code tables. Each feature table stores all thresholds corresponding to the nodes where that feature appears across every tree in the RF. These thresholds are partitioned into ranges, and each range is assigned a unique codeword. The codewords from all feature tables are then combined in a per-tree code table, which maps these combinations to the appropriate leaf nodes, thereby determining the final classification outcome. Even though Planter provides publicly available code of their encoding scheme, their implementation is not use case specific and thus cannot be used out of the box as our encoding solution without heavy modifications. We opted to implement our own adaptation of Planter's RF encoding scheme that is specifically tailored for our engineered features, RF model and RA signaling storm use case. Our encoder takes a trained RF model as input and outputs a set of match action entries that make up the encoded RF model. This model is then added to our controller component that loads into the data plane at runtime.

\noindent \textbf{Traffic Filtering.} The first stage of RAID’s P4 pipeline filters and isolates relevant RA control traffic. Using the P4 programmable parser in the Tofino switch, RAID inspects incoming packets on the F1 interface and parses RRC connection requests. All other types of traffic are forwarded directly to their destination while RRC connection request packets are diverted to the detection pipeline. This filtering reduces unnecessary overhead and ensures subsequent stages only process traffic pertinent to potential overload events while all other traffic remains untouched. 

\noindent \textbf{Pre-processing.} Once RA traffic is filtered, flows are identified by their 5-tuple $\langle$src IP, dst IP, src SCTP port, dst SCTP port, protocol$\rangle$, we compute flow hashes out of the 5-tuples which serve as flow identifiers (Flow IDs). For each flow, the switch maintains a compact feature vector updated per-packet using integer and bit-wise operations. Our features capture both traffic load patterns and short-timescale burstiness that characterize signaling storms. We maintain per-flow state using register arrays where each register slot stores the last seen flow ID. On a hit to an initialized slot whose stored flow ID differs from the packet’s flow ID, we treat this as a hash collision and emit a small digest to the controller to bypass classification for that packet. 

\noindent \textbf{Inference and mitigation.} RAID classifies each flow within the switch pipeline to determine if the flow is malicious or benign. Once a flow’s feature vector is updated, the packet traverses the encoded RF. First, for each feature, the switch finds which threshold range the feature value falls into and assigns it a short code that represents that range. Next, once the encoded RF entries are loaded by the controller into the P4 pipeline, per-tree code tables resolve those codes to a leaf decision. A final voting table aggregates the per-tree outcomes majority vote to produce the final class for the flow. 

Upon classification, if a flow is labeled malicious, we perform in-switch mitigation by dropping that flow through our forwarding table. This approach blocks packets belonging to that flow from reaching the CU’s RRC. Benign flows on the other hand are directly forwarded to the CU. Once a flow is classified, subsequent packets belonging to that flow are directly assigned a class and either forwarded or dropped without going through the inference process again.

For model performance evaluation, monitoring, and closed-loop control, RAID exports a lightweight digest to the controller once a flow is classified and on rare hash-collision events. The digest includes information such as the flow’s 5-tuple, assigned class, packet count and inference duration. The controller also updates the flow tables with rules to skip the inference for classified flows whose class is now known.

\noindent \textbf{Controller.} RAID incorporates an online control-plane component that manages model deployment and runtime configuration while remaining decoupled from the switch’s data plane. The controller performs several key management functions. At P4 runtime, it programs the switch by injecting the pre-encoded RF table entries into the pipeline and setting up port configurations and initial forwarding rules. It includes a digest manager that handles asynchronous digests generated by the data plane, parsing and exporting them for external monitoring and offline analysis. This design ensures that all control and monitoring tasks are handled without interfering with the line-rate inference executed within the data plane.

\section{Experimental Evaluation} \label{sec:experimental_evaluation}\label{sec:evaluation}

\subsection{Setup} \label{subsec:setup}

\noindent\textbf{Testbed.} To evaluate RAID, we deployed a 3GPP-compliant end-to-end 5G standalone (SA) testbed built entirely with Commercial-Off-The-Shelf (COTS) hardware and open-source 5G software similar to the O-RAN architecture shown in Figure \ref{fig:ORAN}. The setup consists of multiple Ubuntu 22.04 servers: one running the Open5GS 5GC \cite{Open5GS}, another hosting the Openairinterface \cite{OpenAirInterface5g} (OAI) 5G CU split into CU-CP and CU-UP, two servers host five OAI DUs each configured for 40 MHz bandwidth with 106 PRBs in Band n78 TDD. The CU and DUs are interconnected through a UfiSpace S9180-32X Tofino-1 switch, where RAID’s P4 pipeline operates inline on the F1 interface between the two. A near-RT RIC running OAI FlexRIC \cite{FlexRIC} is deployed on a separate host. For the radio, we use two Ettus USRP X310 SDRs for real transmission and OAI RFSimulator for radio emulation. For the UEs, we use three Google Pixel 7 Pros and simulate up to 1000 UEs using our custom RA signaling storm traffic generator.

\noindent\textbf{Benchmarks.} A direct performance comparison with other RIC/xApp GPP-based approaches is not possible since, as discussed in Section \ref{subsec:keychallenges} and to the best of our knowledge, RAID is the first RA signaling storm detection and mitigation framework that is specifically designed for disaggregated O-RAN CU/DU architectures and considers realistic multi-DU to single CU layouts where traffic from multiple DUs is aggregated by the same CU.  To mitigate this and ensure a fair comparison, we mainly evaluate RAID against a simulated, GPP-based \texttt{RIC} inference pipeline deployed on our 5G testbed that mimics n-RT RIC and xApp processing to demonstrate the advantages of our in-network inference approach in preventing RA signaling storms while maintaining QoS. To provide further insights on RAID's performance and highlight RAID's scalability on the switch, we deployed a behavioral model version 2 (bmv-2) implementation of RAID, \texttt{BMV2}. bmv2 is P4 programmable switch simulation tool for developing, testing and debugging P4 data planes and control plane software. Albeit useful, it is very limited in terms of throughput and cannot maintain deterministic hardware P4 switch processing latency as it is still deployed on a GPP. We also use the switch’s basic L3 forwarding \texttt{L3\_FWD} as a baseline to highlight the processing overhead introduced by RAID’s in-network ML inference.

\noindent\textbf{Signaling storm simulation.} We use a combination of real and simulated sources of UE traffic. Real-world 5G traffic generated through a physical software defined radio (SDR) and phone UEs connected to our testbed. For simulations, we developed a custom MSG3 traffic generator. Our generator is able to generate up to 50000 MSG3s per second. Given the need for granular control over MSG3 packet profiles and signaling storm characteristics, such as burstiness, transmission intervals, and rate, we implemented the traffic generator using Scapy for precise packet crafting and Python sockets allowing us control over all simulation variables. 

\subsection{Evaluation}

\begin{figure*}
  \centering

  \begin{subfigure}{0.32\textwidth}
\includegraphics[width=\linewidth]{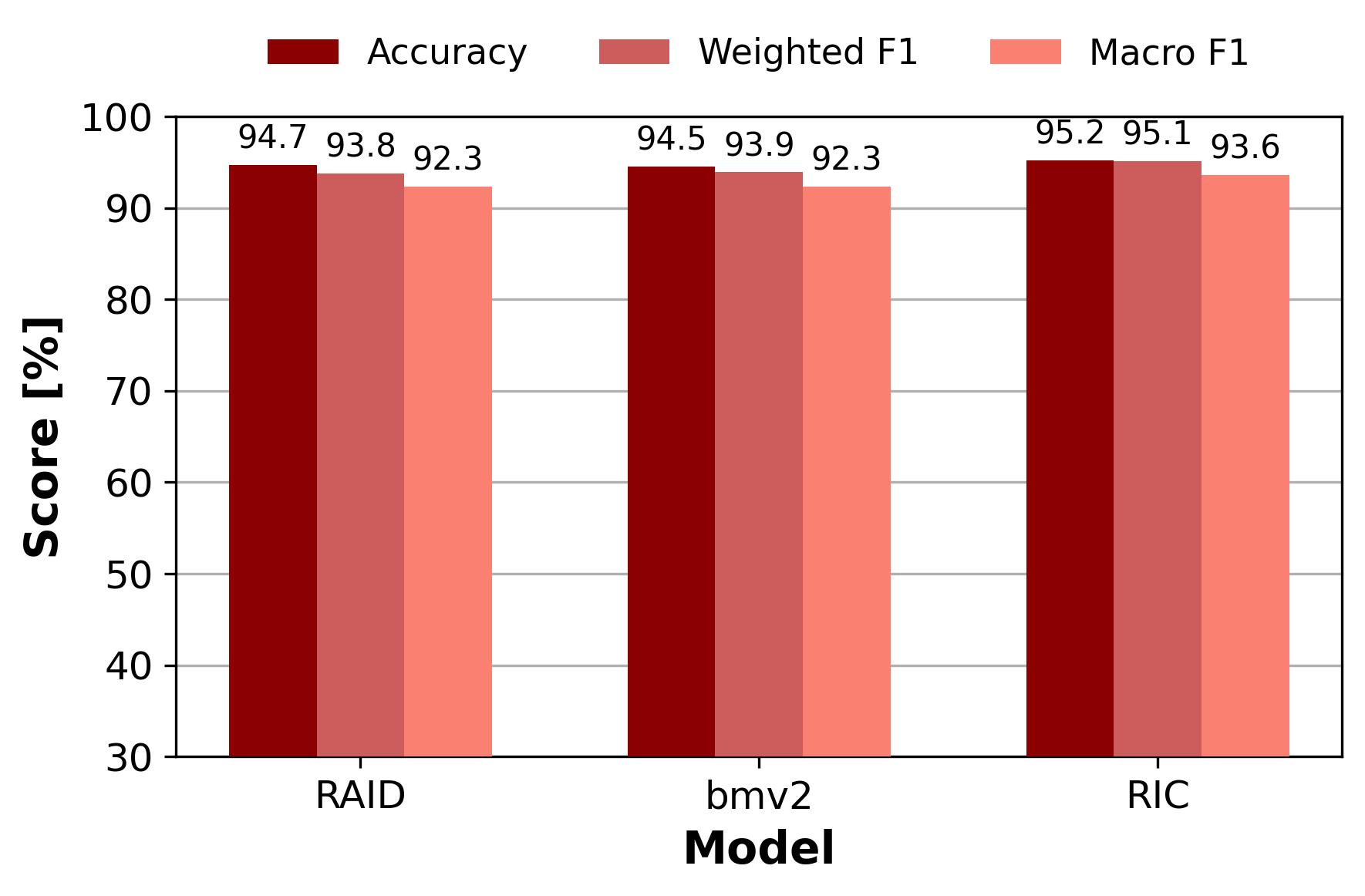}
    \caption{Low Load}
    \label{fig:raid_acc_low}
  \end{subfigure}
  \begin{subfigure}{0.32\textwidth}
\includegraphics[width=\linewidth]{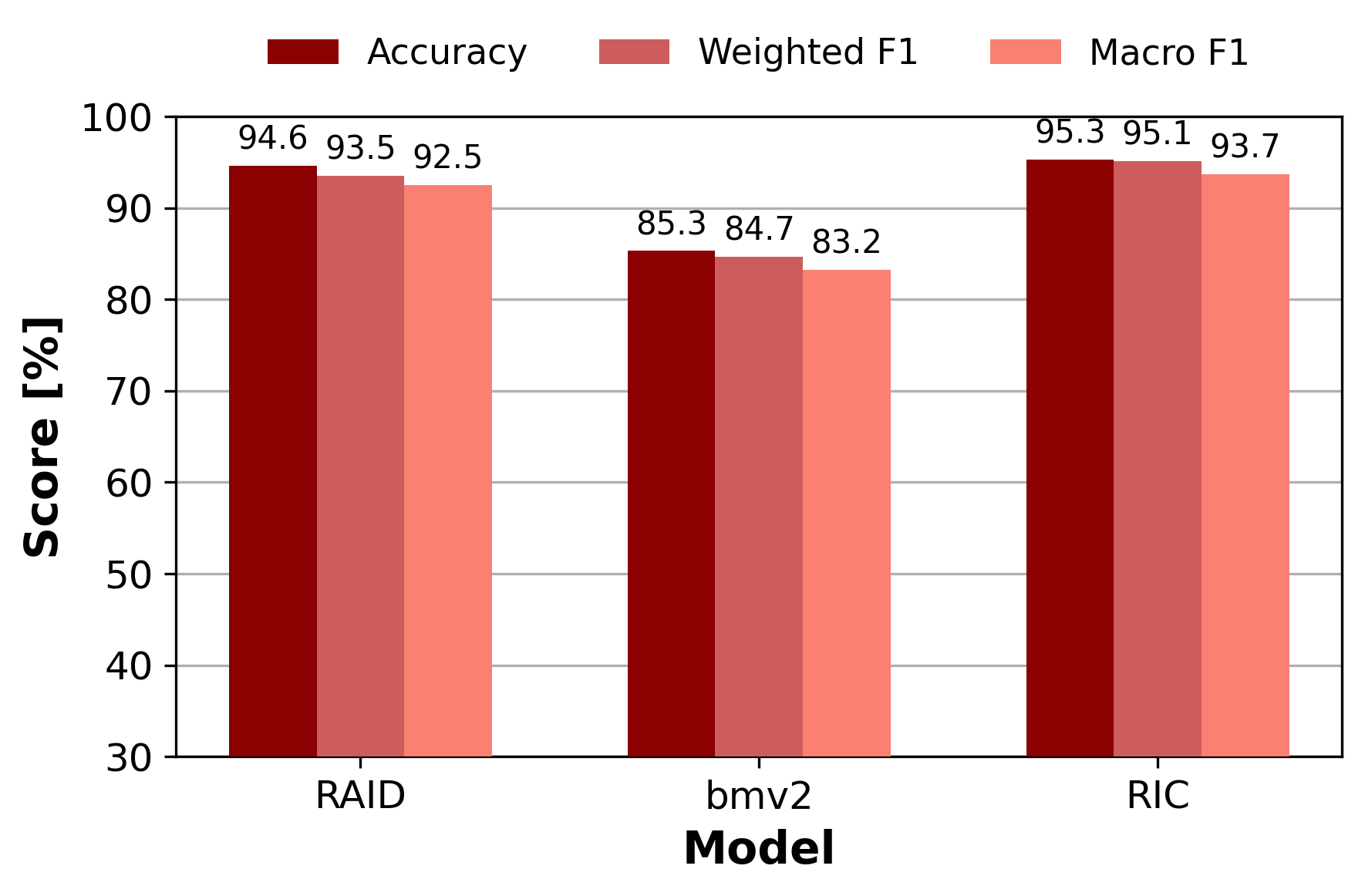}
    \caption{Moderate Load}
    \label{fig:raid_acc_moderate}
  \end{subfigure}  
  \begin{subfigure}{0.32\linewidth}
\includegraphics[width=\textwidth]{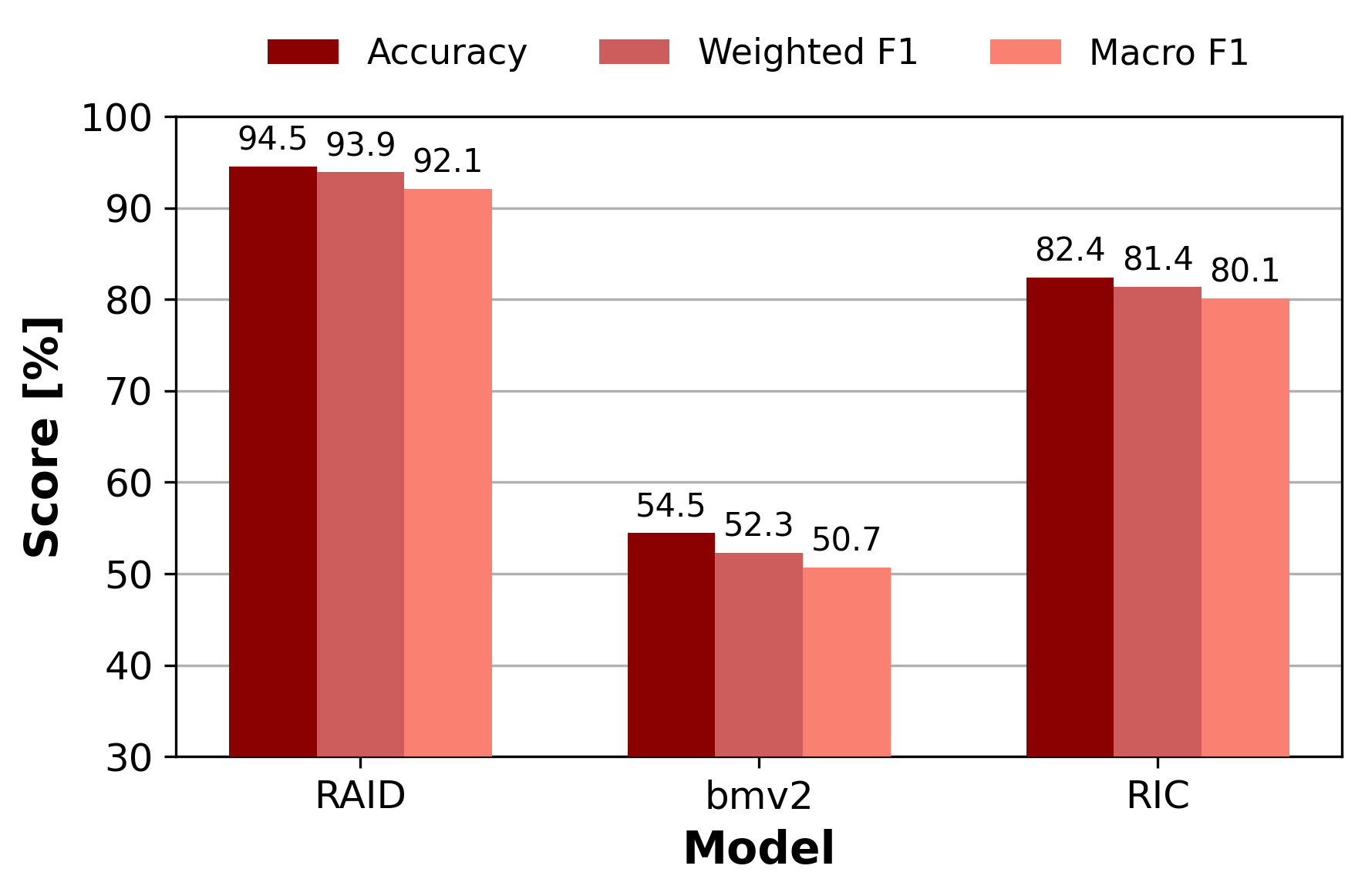}
    \caption{High Load}
    \label{fig:raid_acc_high}
  \end{subfigure}
\caption{RAID classification accuracy}
\label{fig:RAID_ACCURACY}
\end{figure*}

\begin{figure*}
    \centering
   \begin{subfigure}{0.3\textwidth}
\includegraphics[width=\linewidth]{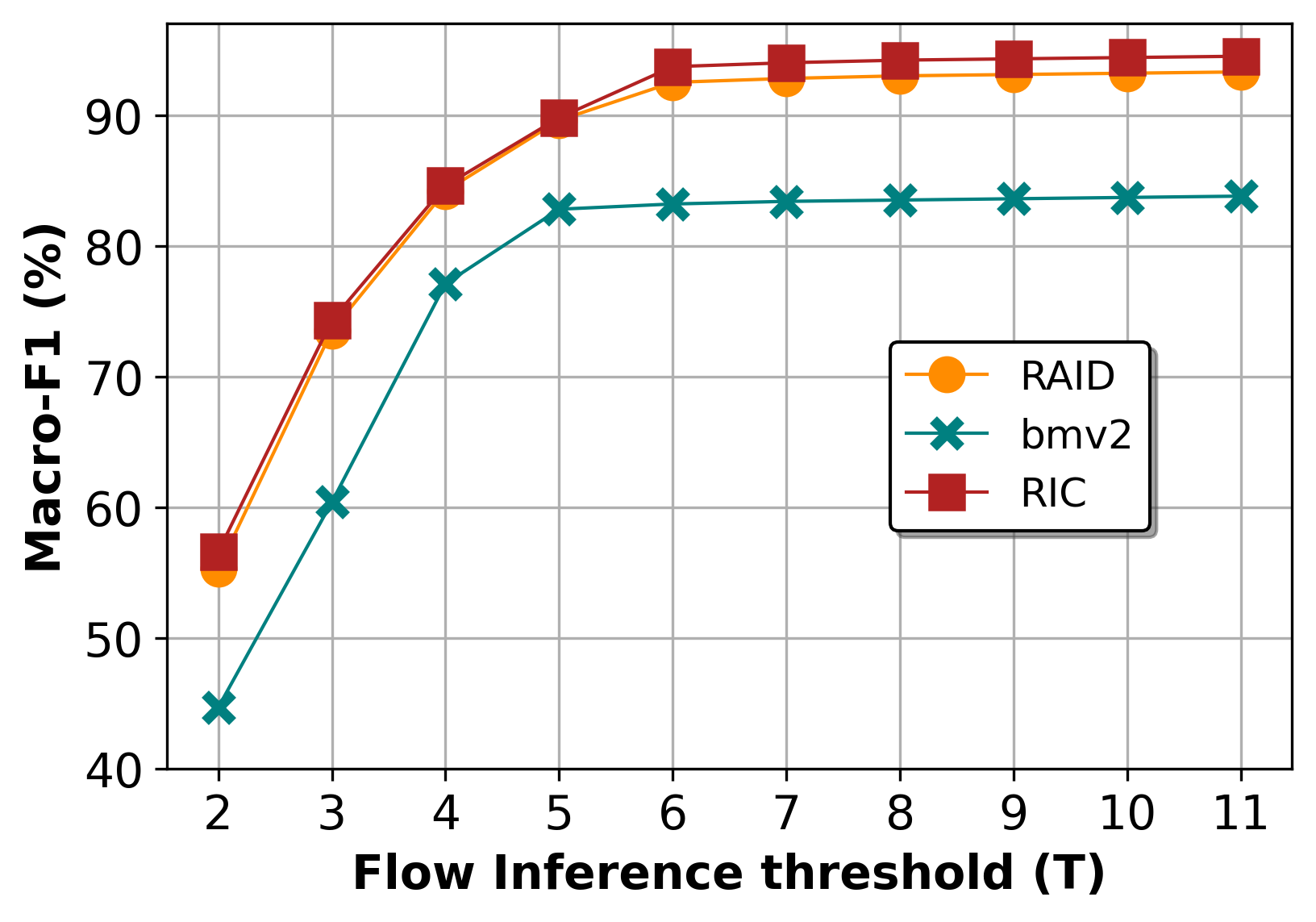}
    \caption{Optimal Inference Threshold}
    \label{fig:raidtvalues}
  \end{subfigure}  
  \begin{subfigure}{0.3\textwidth}
  \scriptsize
  \renewcommand{\arraystretch}{1.8}
    \begin{tabularx}{\linewidth}{XXXX} \toprule
    \textbf{Benchmark}  & \textbf{Low} & \textbf{Moderate} & \textbf{High} \\ 
    \toprule
    \textbf{RAID}  & 3.4 & 3.4 & 3.5    \\ \midrule
    \textbf{bmv2}  & 20.3 & 24343 & 2614434  \\ \midrule
    \textbf{RIC}  & 15340 & 63940 & 176432  \\ \bottomrule
    \end{tabularx}
    \vspace{15pt}
    \caption{Inference Latency (µs)}
    \label{tab:inference-latency}
  \end{subfigure} 
  \begin{subfigure}{0.3\textwidth}
\includegraphics[width=\linewidth]{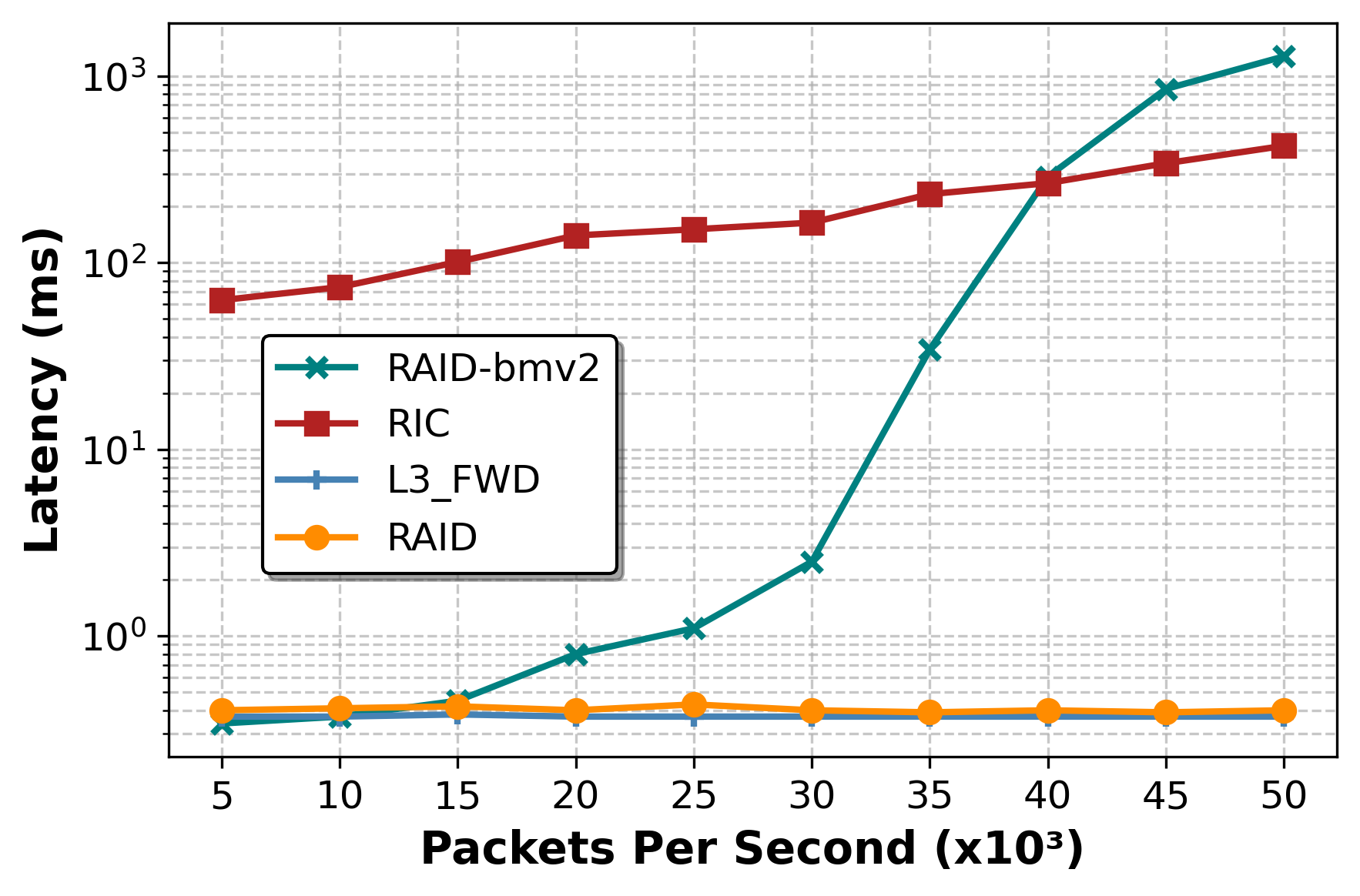}
    \caption{Inference Latency wrt PPS}
    \label{fig:raidppslatency}
  \end{subfigure}   
\caption{RAID Inference performance}
\label{fig:RAID_LATENCY}
\end{figure*}

To ensure statistical significance, each experiment is conducted for 5 minutes and repeated across 25 runs. Results are then averaged to obtain a reliable measure of performance.

\noindent \textbf{Classification accuracy.} We evaluate \texttt{RAID's} classification accuracy using standard ML metrics, notably accuracy and F1-score. These metrics are computed from the confusion matrix counts of true positives (TP), true negatives (TN), false positives (FP), and false negatives (FN). The overall accuracy is defined as

\begin{equation}
\text{Accuracy} = \frac{TP + TN}{TP + FP + TN + FN},
\end{equation}

while the F1-score, which combines recall and precision into a single harmonic mean, is calculated as

\begin{equation}
F_1 = \frac{2TP}{2TP + FP + FN}.
\end{equation}

Figure~\ref{fig:RAID_ACCURACY} presents the classification results across low 10K PPS, moderate 30K PPS, and high 50K PPS traffic loads. Under low load conditions (Fig.~\ref{fig:raid_acc_low}), all three implementations achieve comparable performance, with \texttt{RAID} reaching 94.7\% accuracy, 93.8\% weighted-F1, and 92.3\% macro-F1. The \texttt{bmv2} and \texttt{RIC} setups perform similarly, attaining 94.5\% and 95.2\% accuracy, respectively.  As the load increases to moderate (Fig.~\ref{fig:raid_acc_moderate}), differences between the implementations become more pronounced. \texttt{RAID} maintains strong performance with 94.6\% accuracy, 93.5\% weighted-F1, and 92.5\% macro-F1, closely matching the GPP reference (\texttt{RIC}) which records 95.3\%, 95.1\%, and 93.7\%, respectively. In contrast, \texttt{bmv2} exhibits a sharp degradation to 85.3\% accuracy, 84.7\% weighted-F1, and 83.2\% macro-F1, reflecting the software switch’s limited throughput and its inability to sustain deterministic inference at higher packet rates. The marginal $\sim$1–2\% difference between \texttt{RAID} and \texttt{RIC} confirms that hardware-based deployment retains near-identical accuracy while operating at line rate.
Under high load (Fig.~\ref{fig:raid_acc_high}), the hardware implementation continues to demonstrate stable behavior, maintaining 94.5\% accuracy, 93.9\% weighted-F1, and 92.1\% macro-F1. Meanwhile, \texttt{bmv2} experiences severe degradation, dropping to 54.5\%, 52.3\%, and 50.7\%, respectively. The \texttt{RIC} suffers significant decline, achieving only 82.4\% accuracy and 81.4\% weighted-F1. This divergence shows the scalability advantage of in-switch inference, which sustains accuracy despite the computational and timing pressure of increasing packet rates that eventually saturates GPP resources causing severe QoS degradation. These results confirm that in-network ML classification can achieve near-parity with offline models without compromising precision or robustness under varying network conditions and even scale better at higher loads due to the deterministic nature of P4 pipelines.


    

\noindent \textbf{Inference Threshold.} \texttt{RAID} needs to collect enough flow feature data before triggering the flow inference. To determine the optimal number of packets that must be observed before triggering flow classification, we varied the inference threshold~$T$, defined as the number of initial packets from a flow used to compute the complete feature vector. Since our engineered features depend on flow-level statistics, classification with very small~$T$ values can suffer from incomplete information, while excessively large~$T$ increases detection delay.

Figure~\ref{fig:raidtvalues} shows \texttt{RAID's} Macro-F1 performance as a function of~$T$. The model is unable to infer at the first packet as there is no previously known values to perform feature computations from, notably for the min and max IPD features. As~$T$ increases from~2 to~6, the model is able to infer and the Macro-F1 rises sharply with every incoming packet. Beyond~$T{=}6$, the gain saturates and the curve flattens, showing that larger windows provide diminishing returns.

Let $M(T)$ denote the Macro-F1 score at threshold~$T$. We define the marginal gain as
\begin{equation}
\Delta M(T) = M(T) - M(T-1).
\end{equation}
The optimal threshold $T^\ast$ is chosen as the smallest $T$ satisfying $\Delta M(T) < \varepsilon$, where $\varepsilon$ represents a negligible improvement margin. Experimentally, this condition holds at $T^\ast = 6$.

This achieves the best trade-off between accuracy and latency, \texttt{RAID} reaches a Macro-F1 of approximately~92-93\%, only marginally below the \texttt{RIC}. Before the optimal flow inference threshold $T^\ast$ is hit, a best effort classification is performed on packets 2 to 5 with the available feature information up to that point. The small accuracy gap can be attributed to the integer-based feature encoding required for in-switch inference, which replaces floating-point precision with range-based discretization to meet hardware constraints. The limited per-flow state and bounded register memory on the switch can also slightly reduce feature granularity compared to the GPP-based \texttt{RIC}, which operates with full-precision arithmetic and unconstrained memory. Nonetheless, \texttt{RAID's} accuracy remains near identical while achieving microsecond-scale inference latency, validating the practicality of P4 pipelines.

\noindent\textbf{Inference delay.} We measure the inference delay on a per-flow basis, it represents the time elapsed between the arrival of the first packet of a new flow and the moment the flow is classified once the optimal inference threshold $T^\ast = 6$ is reached. The reported values thus capture the complete inference delay from the initiation of a flow to the generation of its classification decision, averaged across all observed flows under different traffic loads.

\begin{figure*}
  \centering
  \begin{subfigure}{0.32\textwidth}
\includegraphics[width=\linewidth]{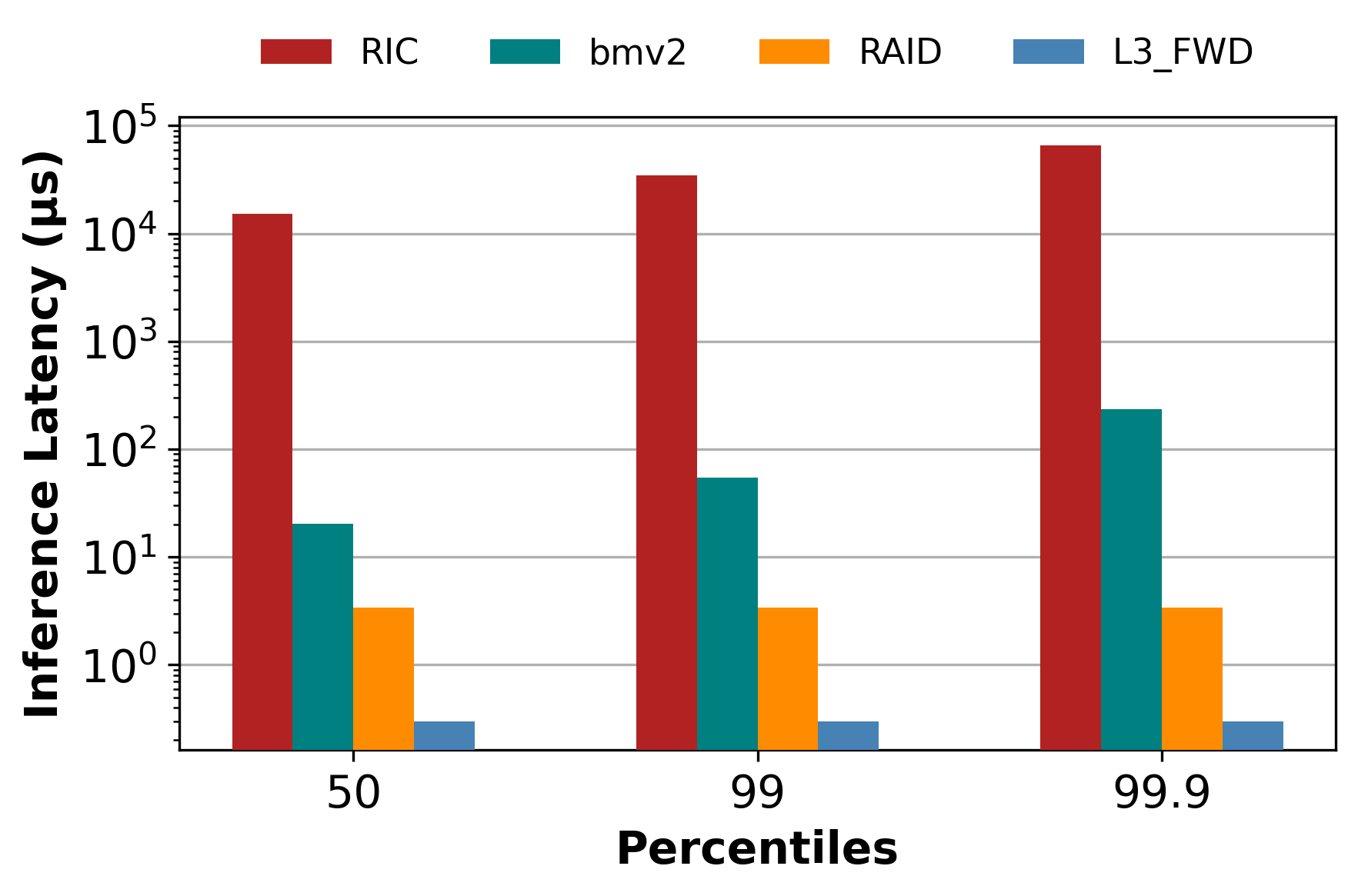}
    \caption{Low Load}
    \label{fig:raid_tail_low}
  \end{subfigure}
  \begin{subfigure}{0.32\textwidth}
\includegraphics[width=\linewidth]{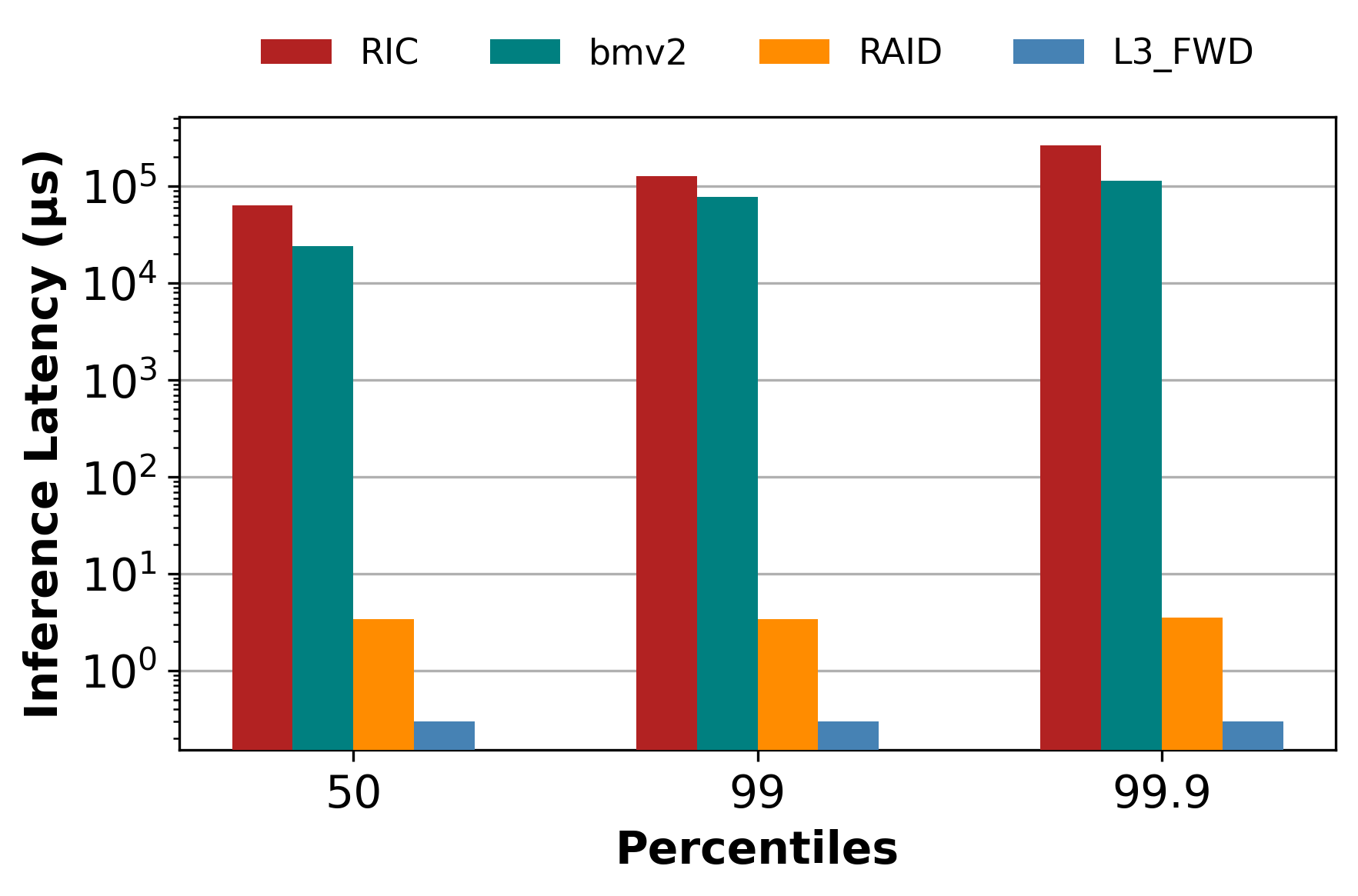}
    \caption{Moderate Load}
    \label{fig:raid_tail_mod}
  \end{subfigure}  
  \begin{subfigure}{0.32\linewidth}
\includegraphics[width=\textwidth]{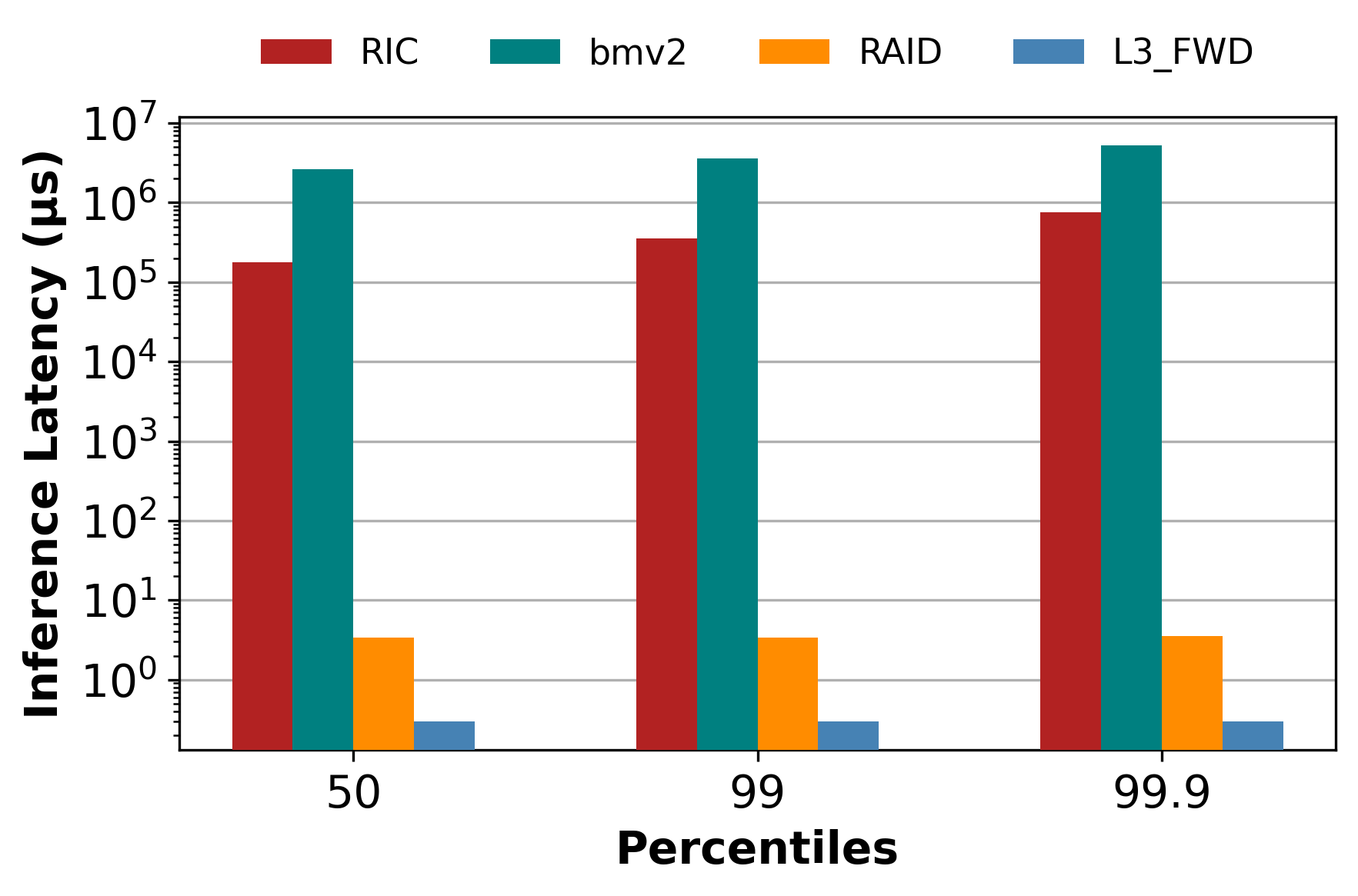}
    \caption{High Load}
    \label{fig:raid_tail_high}
  \end{subfigure}
\caption{RAID Inference tail performance}
\label{fig:RAID_PERCENTILES}
\end{figure*}

Table~\ref{tab:inference-latency} summarizes the median inference delay under varying traffic conditions, while Fig.~\ref{fig:raidppslatency} visualizes median delay changes across increasing packet rates per second and how they compare to the \texttt{L3\_FWD} baseline. \texttt{RAID} achieves and maintains a consistent flow inference delay of approximately $3.4~\mu$s. This stability is attributed the use of a dedicated hardware P4 pipeline which eliminates underlying delays associated with GPP architectures such as kernel processing and context switching operations. The software-based \texttt{bmv2} exhibits rapidly increasing inference delay as packet rate grows rising from $20.3~\mu$s at low load to $24.3$~ms at moderate load and exceeding $2.6$~s under high load. This nonlinear growth highlights that, while \texttt{bmv2} can be a useful tool for low scale testing of in-network ML-based solutions, its unable to handle high, realistic, traffic loads observed during RA signaling storms. The \texttt{RIC} achieves inference delays of $15.3$~ms at low loads, $63.9$~ms at moderate load and $176$~ms under high load. Its reliance on external E2 and control-plane interactions between the CU and the near-RT RIC platform, in addition to GPP and queuing delays makes for a much slower non-deterministic pipeline when compared to \texttt{RAID}. When evaluated against the CRT timing window, \texttt{RAID}'s inference delay remains comfortably below even the tightest bound. By contrast, the \texttt{RIC} exceeds the limit even at moderate load, which is not ideal for latency-critical RA signaling storm detection and mitigation in the RAN.


\noindent \textbf{Tail delay.} Fig.~\ref{fig:RAID_PERCENTILES} illustrates the evolution of inference tail latency distributions across load levels and percentiles. Fig. \ref{fig:raid_tail_low} shows obtained results at low traffic loads. \texttt{RAID} exhibits a deterministic profile with median, 99th, and 99.9th percentile delays virtually identical at $3.4~\mu$s. This invariance is the result of the fixed-cycle nature of the switch pipeline where every packet traverses an identical sequence of M/A stages. In contrast, \texttt{bmv2} and \texttt{RIC} show dispersion in their tails, with the former increasing from $20.3~\mu$s to $2{,}000~\mu$s and the latter from $15.3$~ms to $100$~ms, revealing early onset variability even under light traffic.

As the traffic increases to moderate loads as shown in Fig. \ref{fig:raid_tail_mod} , \texttt{bmv2's} tail diverges sharply from its median, expanding from a $24.3$~ms median to nearly $0.9$~s at the 99.9th percentile, a $\sim3700\%$ increase. Similarly, the \texttt{RIC} grows from $63.9$~ms to $250$~ms at the 99.9th percentile. Under high loads, the contrast becomes extreme. RAID remains stable with only microsecond-level variance between percentiles, while the \texttt{RIC}, experiences heavy-tail growth from $176$~ms to $800$~ms. These widening tails for \texttt{bmv2} and the \texttt{RIC} stem from transient queuing in user-space packet processing that results from congestion, on top of the \texttt{RIC's} control loop delays. Both systems suffer from non-deterministic scheduling which leads to variable processing times leaving \texttt{RAID} as the only viable option to guarantee strict detection requirements imposed by latency-sensitive control applications.

\section{Limitations and Future Work} \label{subsec:limitations}

While RAID demonstrates substantial performance gains over GPP-based RIC/xAPP deployments, it also has certain limitations. The detection and mitigation in RAID's pipeline is based on traffic flows which, as discussed previously is UE identifier and parameter agnostic making it particularly effective at detection malicious signaling storm traffic. The limitation of this approach is that RAID does not have any visibility into which UEs are sending malicious traffic and thus, unable to trace back the origin of the flows beyond the DU. While our simple mitigation of dropping identified attack flows in-switch is effective, malicious UEs are still able to send re-establishment requests as there is no way to identify and blacklist them. For future work, we plan on tackling this shortcoming by introducing a new online control component to RAID that interacts with the P4 pipeline and digest manager with the goal of effectively tracing back identified malicious flows to their original source UE which would allow RAID to blacklist malicious UEs in real-time, making them unable to send any subsequent malicious requests.



\section{Conclusion}\label{sec:conclusion}

RAID demonstrates the feasibility and benefits of offloading O-RAN anomaly detection to programmable data plane hardware. By embedding a ML model into a Tofino switch, RAID enables real-time detection and mitigation of RA signaling storm attacks at line-rate, achieving per-flow inference in mere microseconds. This in-network approach effectively removes the reliance on slow RIC/xApp control loops, allowing the system to identify and mitigate malicious traffic well within tight CRT deadlines of 5G latency-sensitive scenarios. Our empirical results highlight significant performance improvements as RAID maintained high detection accuracy at line-rate even under heavy load, matching the accuracy of a GPP-based RIC solution within a ~1\% margin. We plan to investigate methods for tracing back malicious flow to their origin and generalize the RAID approach to broader classes of O-RAN anomalies by extending the ML models and features to detect other control-plane attacks beyond RA signaling storms.

\section*{Acknowledgments}
This work was supported in part by NSERC Alliance Grant and the Ontario Research Fund – Research Excellence program (Project\# ORF-RE012-051) from the Province of Ontario. The views expressed herein are those of the authors and do not necessarily reflect those of the Province.

\bibliographystyle{IEEEtran}
\bibliography{references}
\end{document}